\documentclass{aastex}
\newcommand{\emaila}{jmould@swin.edu.au}

\begin{document}
\title{Modified Gravity and Large Scale Flows, a Review}
\shorttitle{Large Scale Flows}
\shortauthors{Mould}
 
\author{Jeremy Mould$^{1,2}$}
\altaffiltext{1}{ Centre for Astrophysics \& Supercomputing, Swinburne University of Technology, Victoria 3122, Australia}
\altaffiltext{2}{ARC Centre of Excellence for All-sky Astrophysics (CAASTRO)}

\email{\emaila}

\begin{abstract}
Large scale flows have been a challenging feature of cosmography ever since
galaxy scaling relations came on the scene 40 years ago. The next generation of surveys
will offer a serious test of the standard cosmology. 
\end{abstract}


\section{Introduction}
Peculiar velocities are a probe of large scale structure and the gravitational processes that made it on a larger scale than most other observations. At the turn of the millennium most workers in the field summarized the relationship
between the newly emerged standard cosmology and peculiar velocities as ``nothing to see here.'' Developments since then (Feldman et al 2010, Kashlinsky et al 2011, Magoulas et al 2016, Tully et al 2014) have changed that situation. In this review we pick out the challenges
to $\Lambda$CDM from large scale flows and assess them. We conclude that there
is a very good case for enlarging the data set by an order of magnitude through projects such as Taipan\footnote{www.taipan-survey.org} and DESI\footnote{http://desi.lbl.gov/}, both for statistical reasons (tension
is all very well, but a real challenge to a standard model requires 5$\sigma$ confidence levels) and because the volume so far examined is insufficient.
\begin{figure}
\includegraphics[scale=0.95, angle=-90]{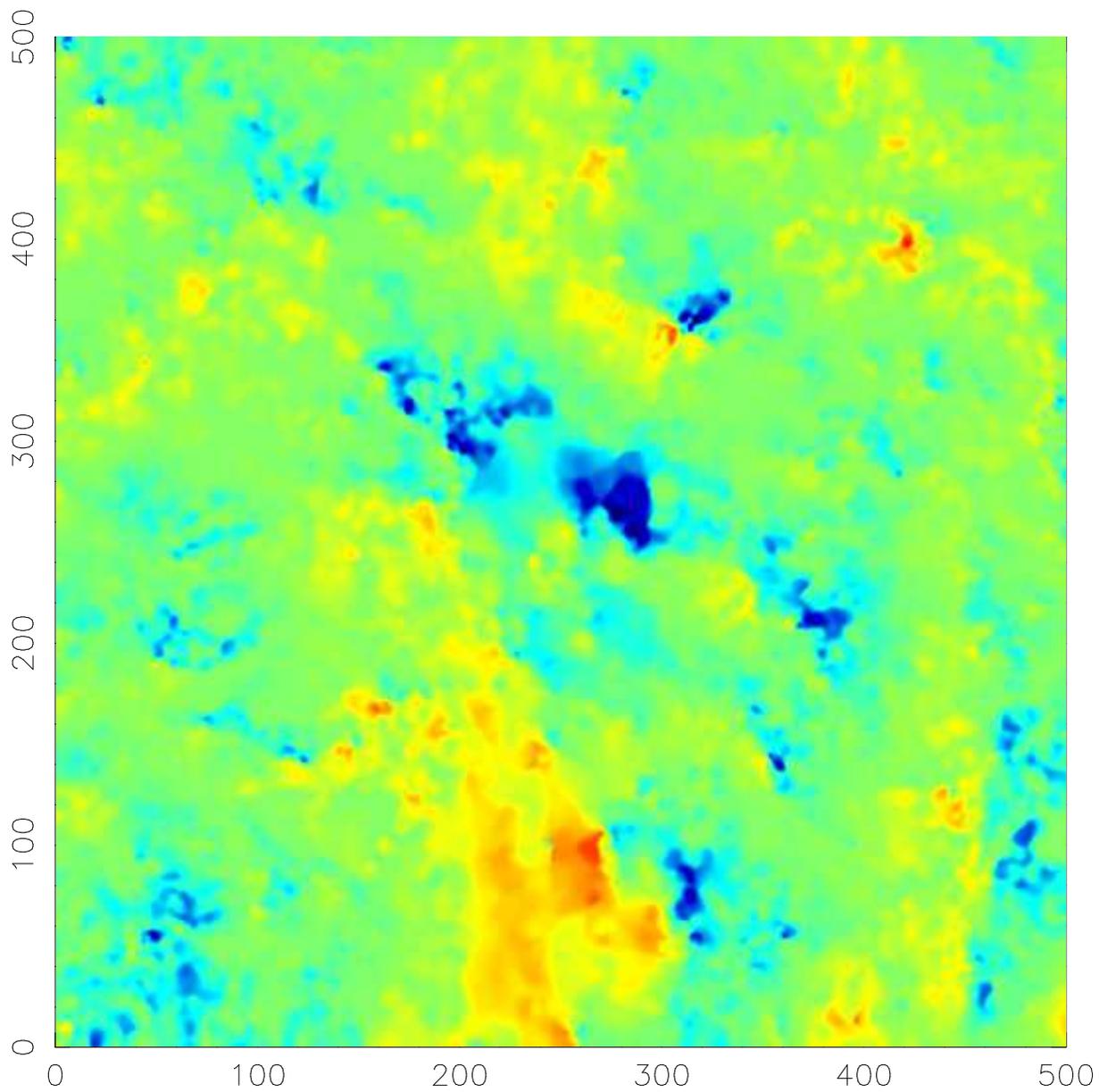}
\caption{Areas of positive (yellow/red) and negative (blue) peculiar velocity as viewed by an observer of the 
Millennium simulation at the origin. The map is a slice through Z = 0 , thickness 10 Mpc. The axes are
labelled in Mpc. Contiguous areas of 100--200 Mpc scale are seen. Data courtesy of the Theoretical Astrophysical Observatory at Swinburne University. }
\end{figure} 
A glance at Figure 1 shows this. A 500 Mpc box from the Millennium simulation shows that the volume containing the zero velocity surface of the largest infalling
regions is not much less than the box itself. Many more simulations are required to achieve the greatest quantitative rigour and to answer the cosmic variance question, ``what is the chance that we are seeing an excessively quiet or excessively disturbed piece of the Universe?"
The velocity field formalism has been reviewed by Ma \& Scott (2014). Improved algorithms are under development (Leclercq et al 2015).

\section{The case for $\Lambda$CDM}
Davis et al (2011) relate the 2MRS galaxy distribution to the mass density field
by a linear bias factor $b$ and include a luminosity dependent, $\propto L^\alpha$, galaxy weighting. They find excellent agreement between observed 
Tully-Fisher relation and predicted galaxy flows, and derive 0.28$~<\beta<~$0.37 
for $\alpha$ = 0. The bias and $\beta$ are related by $\beta~=~f(\Omega)/b$. That $\alpha$ assumption seems unphysical, 
and we compared it with the Millennium simulation with SAGE\footnote{https://github.com/darrencroton/sage} semianalytics (Bernyk et al 2016). 

\begin{figure}
\includegraphics[scale=0.75, angle=-90]{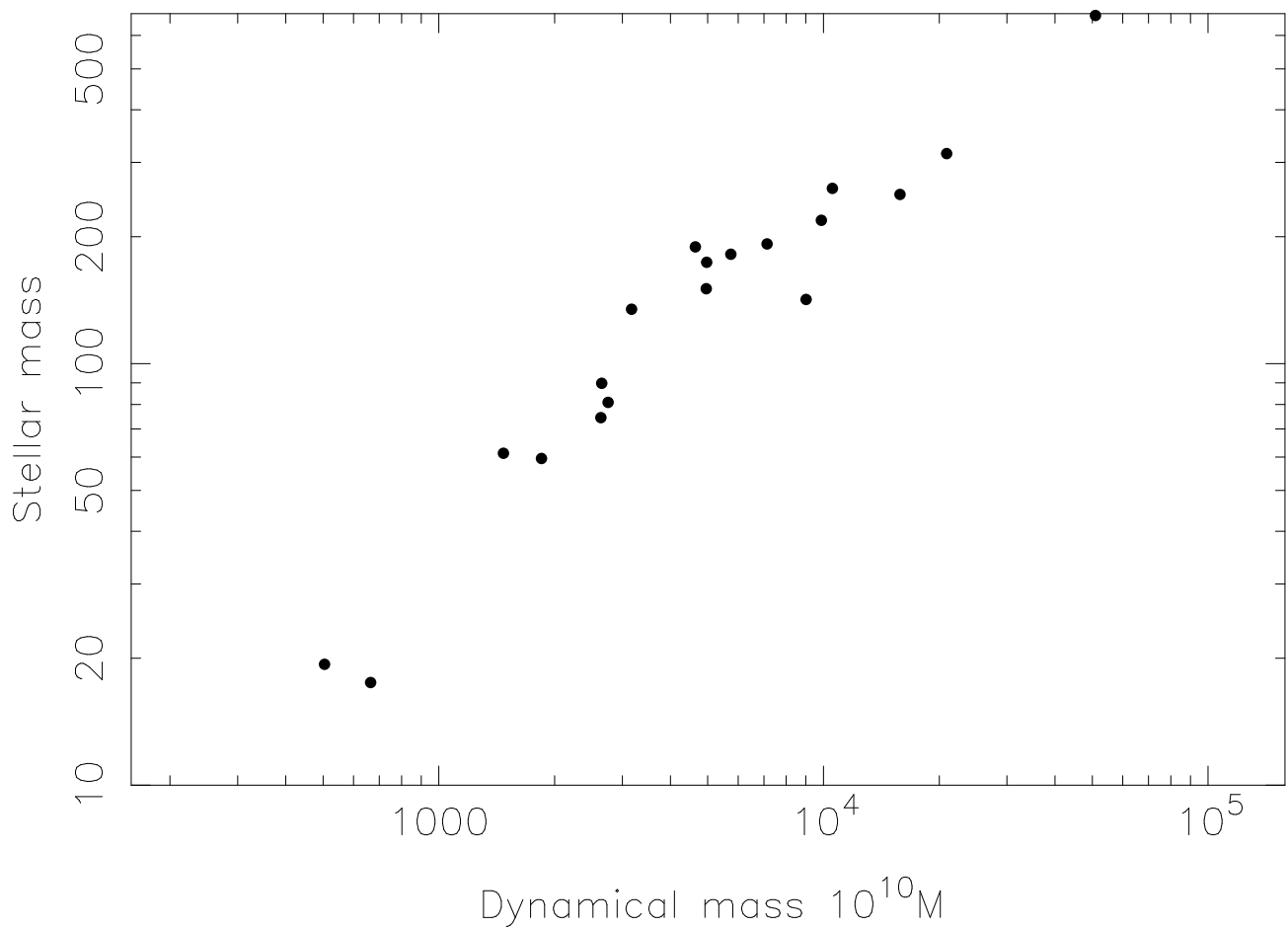}
\caption{The relationship between dynamical mass and stellar mass 
in the Millennium simulation. The horizontal axis is total mass within 10 Mpc spheres; the vertical axis is stellar mass within 10 Mpc spheres.Stellar mass is calculated with the SAGE semianalytic model.
Data courtesy of the Theoretical Astrophysical Observatory at Swinburne University. The units on both axes are 10$^{10}~M_\odot$.}
\end{figure} 

A relation L $\propto~M^{0.7}$ would be a better fit to Figure 2 than $\alpha$ = 0. So would L $\propto~\surd M$, i.e. $\alpha$ = 1/2.
This corresponds to $\beta~\approx$ 0.25 in Figure 12 of Davis et al, but still gives a good $\chi^2$ in their Figure 10.
Springob et al (2016) find similar results with the 2MRS (2MASS redshift survey) Tully-Fisher data.
Tully (2015), using 2MASS K band data found  $\alpha$ = 1.15.%

The latest addition to the Cosmic Flows 3 catalog (Tully et al. 2016) is the
fundamental plane (FP) peculiar velocities of the 6dF galaxy survey (Campbell et al 2015). 
Figure 3 shows that the FP is based on a linear relation between halo mass and virial mass 
after the former has been obtained from stellar mass using the halo occupancy distribution (HOD) 
methodology (e.g. Moster et al 2010). This tends to resolve the FP tilt problem discussed 
by Mould (2014), as the remaining departures from a 1:1 relation may be attributed to a bottom heavy IMF in
large $\sigma$ galaxies (see Lagattuta et al 2016 and references therein). 
Ultimately we trust the empirical calibration of the FP over a theoretical calibration, but we do need to advance beyond 
a pure empiricist view that there is some kind of halo/bulge/disk conspiracy, rather than a theory that is greatly improved by the current generation of simulations, but still incompletely understood.

\begin{figure}
\includegraphics[scale=0.75, angle=-90]{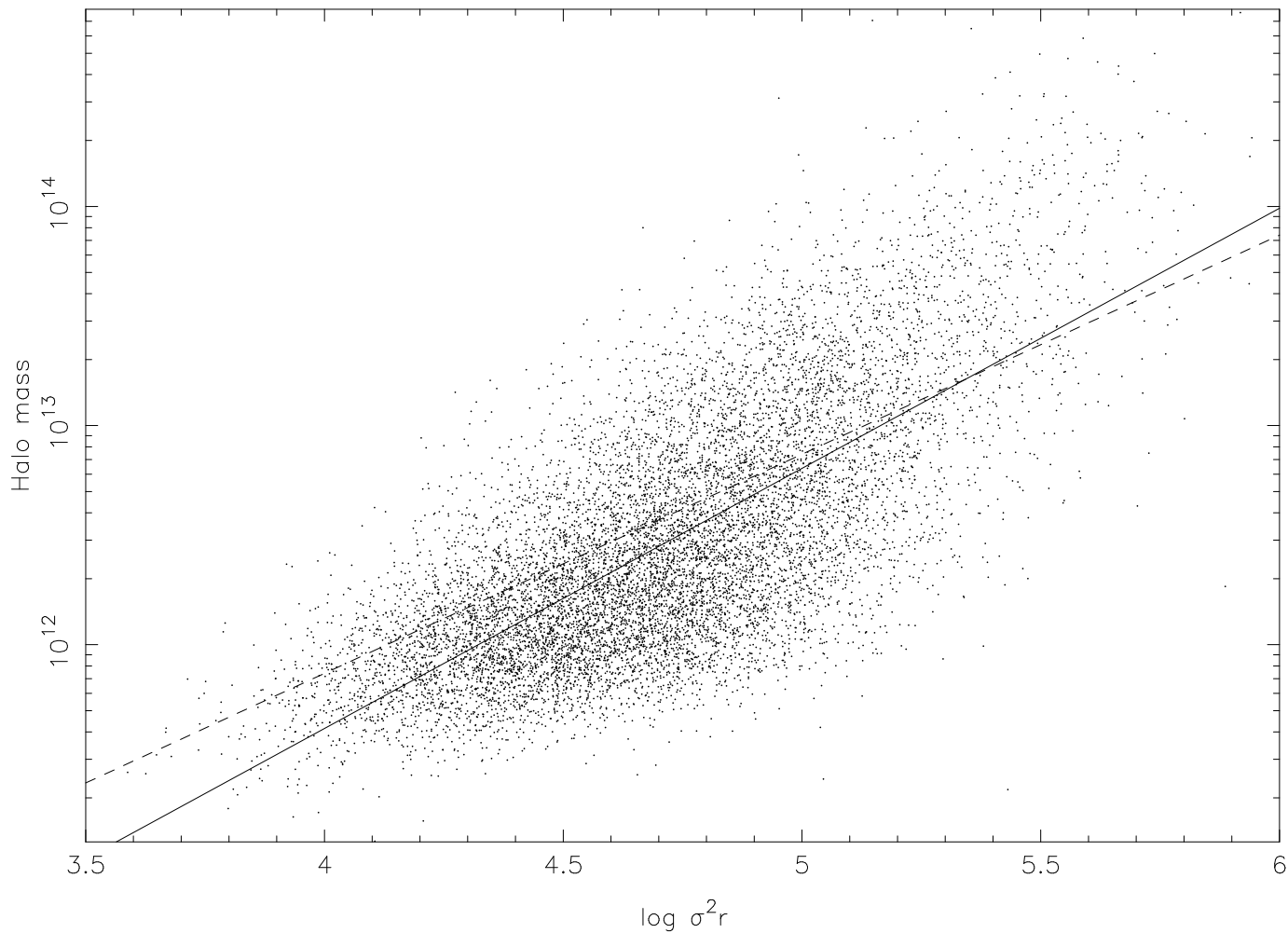}
\caption{Halo mass versus 6dFGS virial mass. The two lines are a least squares fit and a (dashed)
line of unit slope. The halo mass has been obtained from the stellar mass following the HOD formalism of 
Moster et al (2010). A Salpeter IMF was adopted. The least squares fit was a regression minimizing
halo mass errors only.}
\end{figure} 

Johnson et al (2014) show that galaxy flows from 6dFGS\footnote{ http://www.6dfgs.net (Jones et al 2009).} are consistent with the
expectations of $\Lambda$CDM, except possibly at the largest scales, where there
appears to be a 2$\sigma$ positive deviation in the data. Comparison of power spectra, however, loses the phase information in the velocity field, whereas a measurement of coherence length\footnote{H. Feldman presented velocity correlation tensor data at the 2016 Large Scale Flows meeting at ICISE.}  might pose a tighter constraint. A gross mismatch in coherence length was sufficient to rule out
MOND from 6dFGS (Mould et al 2015).

A more severe challenge to $\Lambda$CDM comes from the ``dark flow'' detected by
Kashlinsky et al (2015) by means of the kinematic SZ effect.
At 600--1000 km/s on scales of $\sim$500--1000 Mpc this is clearly incompatible with the standard model. However, such a large dark flow may also be incompatible with the well measured and modest anisotropies in the cosmic microwave background (Ade et al 2014). 
Kinematic SZ studies are ongoing, as the dark flow analysis techniques have been challenged.

\section{The known unknowns}
Reconstructions of the density field from redshift surveys differ and this causes worrying uncertainties in comparisons of peculiar velocity predictions with data.
\begin{figure}
\includegraphics[scale=0.75, angle=-90]{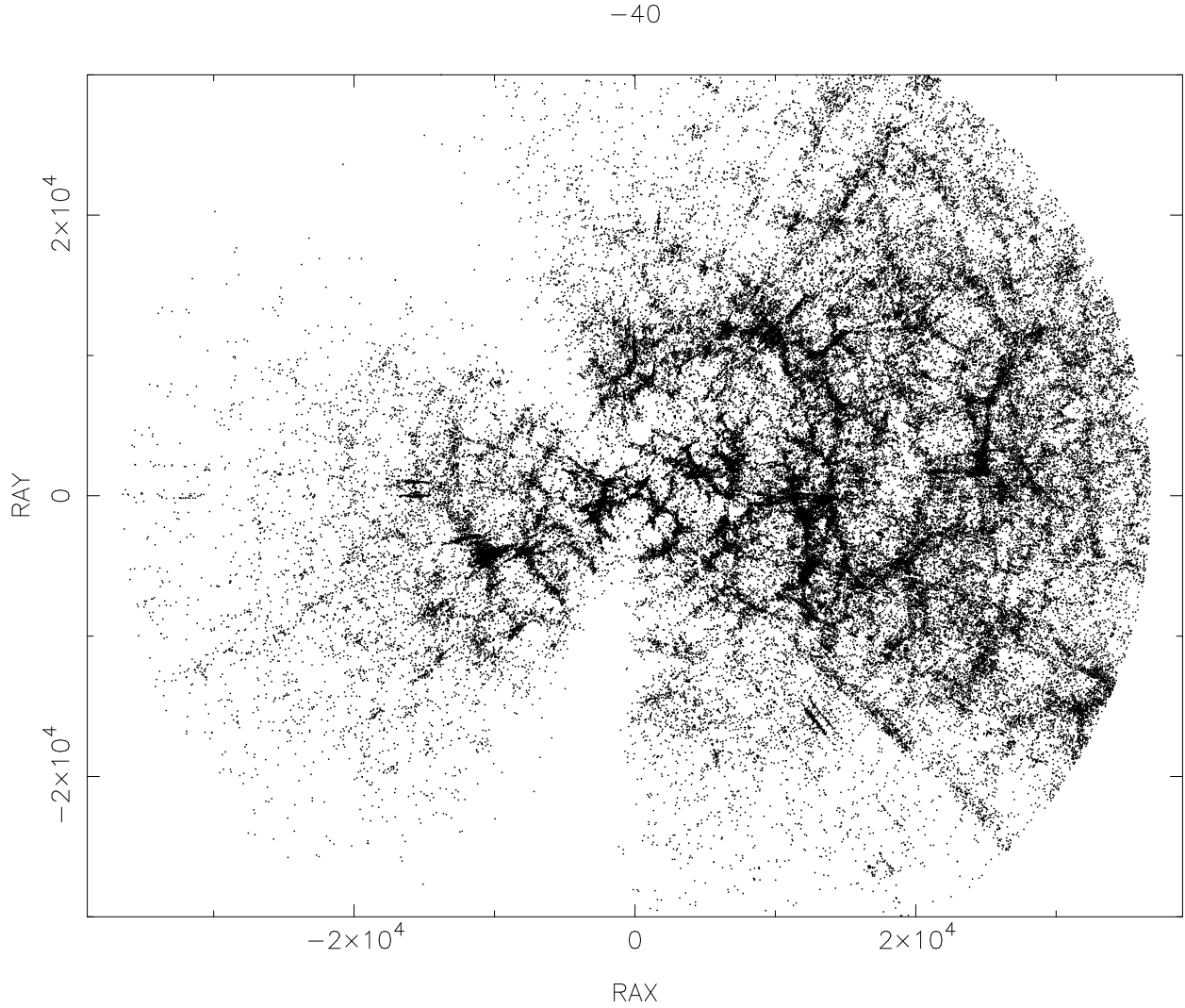}
\caption{A galaxy redshift map from NED at declination --40$^\circ$. The units on the RA axes are km/s. The completeness is better on the SGP side of the sky than the galactic plane side. Data courtesy Lucas Macri.}
\end{figure} 
This is an opportunity for the next generation of local redshift surveys (see Figure 4.) It is essential that such surveys be rigorously defined, so that completeness is uniform. Important work is being done at low galactic latitudes by Said et al (2016) and 
Kraan Korteweg et al (2017).
It is straightforward to calculate the significance of improved reconstructions. For example, multiplying the M/L of the Shapley Supercluster in 2MRS by a factor of a few doubles the predicted local motion in that direction.

The Taipan survey with its goal of a million local redshifts is the next generation southern hemisphere survey. Returning to Figure 1, we see that cosmic variance and the error budget for peculiar velocities demands an equal effort in the northern hemisphere (NH). We can build on the nearly $\pi$ sr of NH coverage in SDSS. Outside this region of the NH the LAMOST survey (Kong \& Su 2010) allots some fibres to galaxies and achieves
good S/N for velocity dispersion measurement. But the details of how appropriate NH coverage will be accomplished are not yet finalized.

\section{A strawman: f(R) gravity}
The current paradigm is to fit the data to the $\Lambda$CDM model with
parameters $\Omega_m$, the bias factor, $b$, and $\sigma_8$. However, if the
flow field fits the model well on 3 Mpc scales and poorly on 30 Mpc scales,
alternative models deserve attention. In the context of General Relativity
there is almost unlimited scope for modified gravity theories (Joyce et al 2016). One class of the
theories that modifies the Einstein tensor $G_{\mu\nu}$ is f(R) gravity (Carroll et al 2004). This modifies the action from the standard Einstein-Hilbert action
to be some new function of the Ricci scalar R. Seiler \& Parkinson (2016)
use 6dFGS to put an upper bound on f$_{R0}$ of 10$^{-4}$ which in simulations gives a bulk
flow of 283 $\pm$ 18 km/s in a sphere of radius 30 Mpc. The parameter
f$_{R0}$ $\rightarrow$ 0 as one approaches $\Lambda$CDM.

In Seiler \& Parkinson's f(R) gravity the Poisson equation for the 
gravitational potential $\Phi$ deviates somewhat from the GR version,
$\delta R(f_R)$ = --8$\pi G \delta\rho_M$, where $\rho_M$ is the density of matter,
and this requires a Poisson solver for exact computation. However,
in underdense regions $\delta R(f_R)$ vanishes and gravity is simply enhanced
by a factor of 4/3. In overdense regions the GR Poisson equation returns.
If we switch between these two limits at $\delta\rho_M/\rho_M$ = 0, we have asymptotic version of this f(R) gravity without the f$_{R0}$ parameter. This is compared
with regular GR\footnote{Based on the density field of Erdo{\u g}du (2006).} in Figure 5.
\begin{figure}
\includegraphics[scale=0.75, angle=-90]{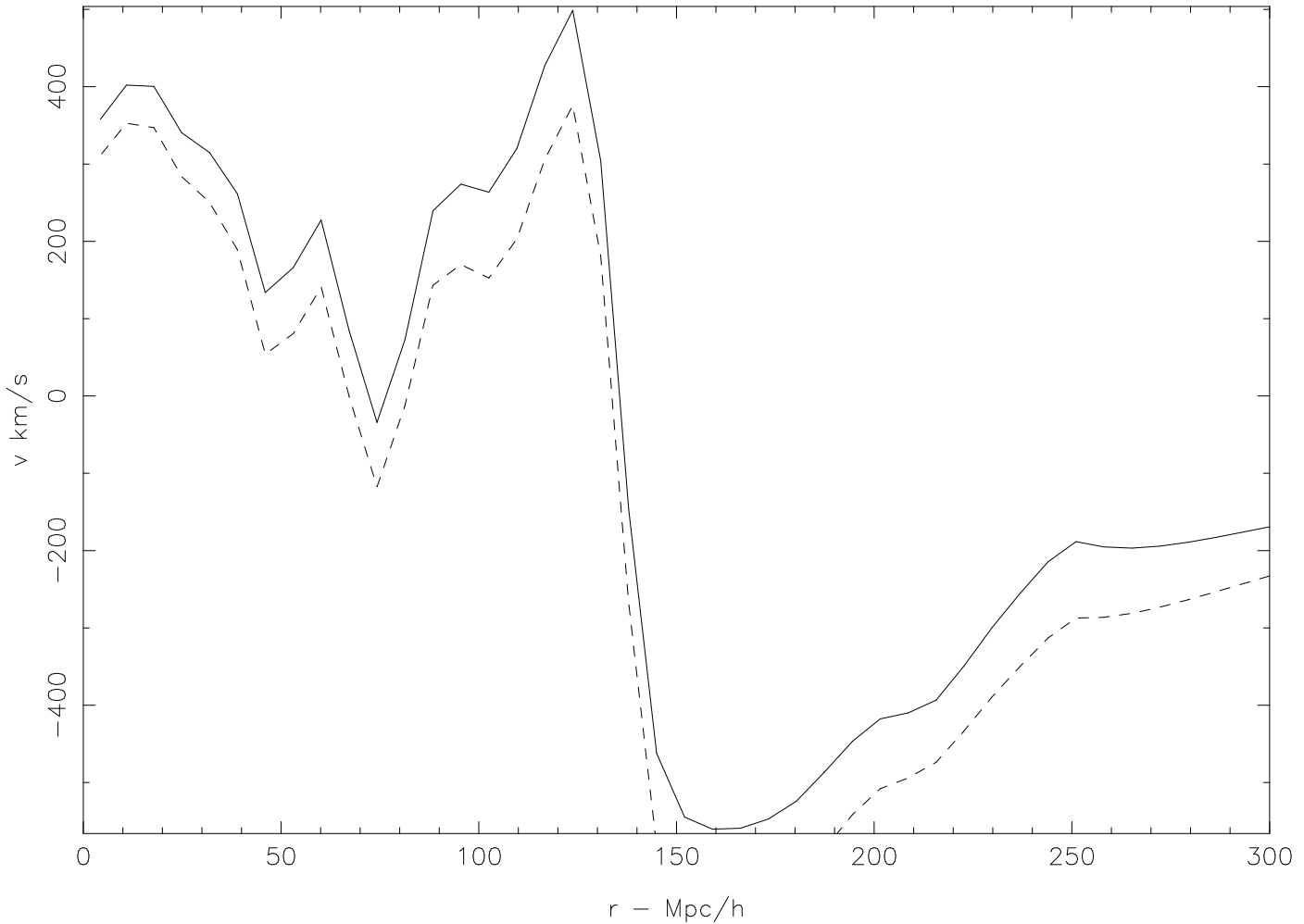}
\caption{Predicted peculiar velocity along the radial diagonal of the second quadrant of the supergalactic plane. The solid curve is our asymptotic version of f(R) gravity; the dashed curve is regular GR. Modified gravity increases the
bulk flow towards the Great Attractor (GA, Dressler 1991) observed from  our perspective. The GA causes the first peak at the left and Shapley the second peak at 120 Mpc/h.
Beyond the GA the prediction drops rapidly (not something which has been observed) and then
rises steadily towards the Shapley Supercluster.}
\end{figure} 

As the data improve, modified gravity models become worthy of comparison 
with the standard model. A useful strawman for this purpose is f(R) gravity.

\section{Conclusions}
Peculiar velocities map current epoch cosmic structure on the largest scales.
Following the early excitement of the Great Attractor, all sky surveys
have tended towards consistency with the standard cosmology. New data this decade have
increased the tension, however, and there is now a strong case for collecting
an order of magnitude more high quality data to test $\Lambda$CDM
in an independent way, which did not feature in the Dark Energy Task Force roadmap (Albrecht et al 2006).

In parallel with improving the well calibrated database of redshift independent distances it is necessary to improve the local density map\footnote{For an up to date 3D model see https://skfb.ly/Iy7R by M. Hudson.}
, so that we can be highly confident of our reconstructions and velocity field predictions. Challenges include inevitable 
incompleteness in the Zone of Avoidance, sample inhomogeneity and other observational effects.
Only when these challenges are met, will modified gravity models move from an interesting
option to something that may be required.

\section*{Acknowledgements}
Participants in the recent Large Scale Flows conference would like to thank Roland Triay and ICISE in Vietnam.
I acknowledge NED\footnote {https://ned.ipac.caltech.edu/} and thank Lucas Macri and Pirin Erdo{\u g}du for making data available.
Parts of this research were conducted by the Australian Research Council Centre of Excellence for All-sky Astrophysics (CAASTRO), through project number CE110001020.
Data used in this work was generated using Swinburne University's Theoretical Astrophysical Observatory (TAO). TAO is part of the Australian All-Sky Virtual Observatory (ASVO) and is freely accessible at https://tao.asvo.org.au.
The Millennium Simulation was carried out by the Virgo Supercomputing Consortium at the Computing Centre of the Max Planck Society in Garching. It is publicly available at http://www.mpa-garching.mpg.de/Millennium/.
The Semi-Analytic Galaxy Evolution (SAGE) model used in this work is a publicly available codebase that runs on the dark matter halo trees of a cosmological N-body simulation. It is available for download at
https://github.com/darrencroton/sage.

\section*{References}
\noindent  Ade, P. et al 2014, A\&A, 561, 97\\
Albrecht, A. et al 2006, astro-ph/0609591\\
Bernyk, M. et al 2014, ApJS, 223, 9\\
Erdo{\u g}du, P. et al 2006, MNRAS, 373, 45\\
Campbell, L. et al 2014, MNRAS, 443, 1231\\
Carroll, S. et al 2004, PhysRevD, 70, 043528\\
Davis, M. et al 2011, MNRAS, 413, 2906\\
Dressler, A. 1991, ASP Conf Series, 15, 19\\
Feldman, H. et al 2010, MNRAS, 407, 2328\\
Johnson, A et al 2014, MNRAS, 444, 3926 \\
Jones, DH et al 2009, MNRAS, 399, 683\\
Joyce, A. et al 2016, ARNPS, 66, 95\\
Kashlinsky, A. et al 2011, ApJ, 732, 1\\ 
Kong, X. \& Su, S. 2010, IAUS, 262, 295\\	
Kraan-Korteweg, R. et al 2017, MNRAS, 466, L29\\	
Lagattuta, D. et al 2016, ApJ, in press\\
Leclercq, F. et al 2015, arXiv, 1512.02242	\\
Ma, Yin-Zhe \& Scott, D. 2014,	Astronomy \& Geophysics, Volume 55, 3, 3.33\\
Magoulas, C et al 2016,  IAUS, 308, 336\\
Moster, B. et al 2010, ApJ, 710, 903\\
Mould, J. 2014, astro-ph 1403.1623\\
Mould, J. et al 2015, ApSS, 357, 162\\
Said, K. et al 2016, MNRAS.462, 3386\\
Seiler, J. \& Parkinson, D. 2016, MNRAS, 462, 75\\
Springob, C. et al 2016, MNRAS, 456, 1886\\
Tully, R. B. et al 2014, Nature, 513, 71\\
Tully, R. B. 2015, AJ, 149 ,171\\
Tully, R. B. et al 2016, AJ, 152, 50\\

\end{document}